\documentclass[useAMS,a4paper,usenatbib]{mn2e}
\usepackage{graphicx} 
\usepackage{multirow}
\usepackage{array}
\usepackage{rotating}
\usepackage{color}

\makeatletter
\def\compoundrel#1\over#2{\mathpalette\compoundreL{{#1}\over{#2}}}
\def\compoundreL#1#2{\compoundREL#1#2}
\def\compoundREL#1#2\over#3{\mathrel
     {\vcenter{\hbox{$\m@th\buildrel{#1#2}\over{#1#3}$}}}}
\makeatother

\title[Disruption of Satellites]{Tidal disruption of satellites and formation of narrow rings}

\author[Z. M. Leinhardt et al.]{Zo\"e M. Leinhardt$^{1,2}$\thanks{E-mail:Zoe.Leinhardt@bristol.ac.uk}, Gordon I. Ogilvie$^{1}$, Henrik N. Latter$^{1}$, and Eiichiro Kokubo$^{3}$\\
$^{1}$Department of Applied Mathematics and Theoretical Physics, University of Cambridge, Cambridge CB3 0WA\\
$^{2}$School of Physics, University of Bristol, BS8 1TL\\
$^{3}$National Astronomical Observatory of Japan, 2-21-1 Osawa, Mitaka, Tokyo 181-8588, Japan}

\begin{document} 
\date{Accepted 2012 May 16. Received 2012 May 16; in original form 2012 March 28}
\maketitle

\begin{abstract}
In this paper we investigate the formation of narrow planetary rings
 such as those found around Uranus and Saturn through the tidal
 disruption of a weak, gravitationally bound satellite that migrates
 within its Roche limit. Using $N$-body simulations, we study the
 behaviour of rubble piles placed on circular orbits at different
 distances from a central planet. We consider both homogeneous
 satellites and differentiated bodies containing a denser core. We
 show that the Roche limit for a rubble pile is closer to the planet
 than for a fluid body of the same mean density. The Roche limit for
 a differentiated body is also closer to the planet than for a
 homogeneous satellite of the same mean density. Within its Roche
 limit, a homogeneous satellite totally disrupts and forms a narrow
 ring. The initial stages of the disruption are similar to the
 evolution of a viscous fluid ellipsoid, which can be computed
 semi-analytically. On the other hand, when a differentiated 
 satellite is just within the Roche limit only the mantle is disrupted.
 This process is similar to Roche-lobe overflow in interacting binary
 stars and produces two narrow rings on either side of a remnant
 satellite. We argue that the Uranian rings, and possibly their
 shepherd satellites, could have been formed through the tidal
 disruption of a number of protomoons that were formed inside
 the corotation radius of Uranus and migrated slowly inwards as a
 result of tidal interaction with the planet.
 \end{abstract}

\section{Introduction}

This paper is the first in a series investigating the formation and
evolution of planetary rings. The outer planets of the solar system
harbour ring systems of surprising diversity. Although they have been
explored through space missions as well as ground-based observations,
and Saturn's rings are currently being imaged with unprecedented
resolution by the \textit{Cassini} spacecraft, fundamental questions
still remain about the formation and evolution of planetary rings.

There are four main hypotheses for the formation of the ring systems:
(i) the rings are primordial and were formed directly from the
circumplanetary disk \citep{Pollack:1975,Charnoz:2009kx}; (ii) the rings were formed subsequently from a
satellite that was disrupted by an impact within the Roche limit of
the planet \citep{Harris:1984,Charnoz:2009}; (iii) the rings were formed from the tidal disruption of a
satellite that migrated within the Roche limit \citep{Canup:2010}; (iv) the rings were formed by the tidal disruption of a passing comet \citep{Dones:1991}. Of course it is
possible that more than one formation mechanism is responsible for the
various ring systems.

The study of planetary rings is important not only to explain their
remarkable structure today but also because the history of the rings
may be intimately connected to the formation and evolution of the
planet. If the ring systems can be interpreted correctly, a
considerable amount may be learned about their host planets. In
addition, planetary rings give us the opportunity to make a detailed
study of the behaviour of astrophysical disks and their interaction
with satellites, processes that are important in settings that range
from protoplanetary systems to galactic nuclei.

In this series of papers we will focus on the narrow, dense rings of
Uranus, although similar rings are also found at certain locations
around Saturn (including the F ring, although its proximity to the
Roche limit may make it a separate case). The present paper is
concerned with the formation of narrow rings, while subsequent work
will treat their dynamics and evolution. Narrow, dense rings have
received relatively little attention and offer many theoretical
puzzles. They are the main constituents of the Uranian ring system,
where they are usually assumed to be shepherded by small satellites. Narrow
rings are typically eccentric and inclined, and may contain other
nonlinear oscillation modes \citep{Porco:1990}. Questions abound concerning
the formation of narrow rings, their present confinement by shepherd
satellites or other mechanisms, the assembly and evolution of
shepherding configurations, the lifetime of narrow rings, the origin
of their eccentricity, inclination and other oscillation modes, and so
on.

The Uranian system has 13 known rings \citep{Showalter:2006}, three of
which ($\zeta$, $\mu$ and $\nu$) are broad dust rings and will not be
considered here. The low geometric albedo of the Uranian rings, which
makes them difficult to detect, indicates that their composition
includes at least some dark material \citep{Ockert:1987,
 Karkoschka:1997}. Most of the narrow rings themselves are effectively devoid of
dust particles, having particle radii between 0.1 and 10~m \citep{Esposito:1991}. The $\lambda$ ring is the exception and is composed of sub-$\mu \rm{m}$ and $\mu \rm{m}$-sized dust particles \citep{Showalter:1993, Burns:2001}.
 The rings are either optically thick or close to it
\citep[e.g.][]{Karkoschka:2001} and very narrow with typical widths of
a few km. It is assumed that the lack of small dust particles is due
to aerodynamic gas drag from the extended Uranian exosphere \citep{Broadfoot:1986}.

There are many questions about the Uranian ring system that remain to
be answered: (i) How did the rings form? (ii) How old are the rings?
(iii) How are the rings' narrow width and sharp edges maintained? It
is possible that all the narrow rings (6, 5, 4, $\alpha$, $\beta$,
$\eta$, $\gamma$, $\delta$, $\lambda$, and $\epsilon$) are shepherded
by small satellites. However, only two potential shepherds (Cordelia
and Ophelia) have been detected, located on either side of the
$\epsilon$ ring.  We will address the spreading and confinement of
narrow rings in future work.

The most accepted formation model for the Uranian rings is a series of
disruptive collisions of pre-existing satellites
\citep{Esposito:2006}. However, this model does not explain why most of
the rings are interior to the known satellites. In this paper
we suggest that it is more likely that satellites were tidally
disrupted as they migrated within the Roche limit. Tidal disruption
has also been suggested for the formation of the Saturnian ring
system.  \citet{Canup:2010} proposed that the partial mantle
disruption of a large Titan-sized satellite could explain many
properties of the rings including their composition of nearly pure
water ice.  The gas disk surrounding Saturn early in its history
causes the satellite to migrate within its Roche limit and be
partially disrupted, with the remnant satellite and a significant
fraction of the ring material being consumed by Saturn. Here we
suggest that each narrow ring in the Uranian system could have
resulted from the complete or partial tidal disruption of a single
small satellite.  Unlike \citet{Canup:2010}, we do not require the
tidal disruption to occur early in the history of Uranus when a gas
disk is present. In the Uranian system the corotation radius is well
outside the Roche limit for a homogenous fluid satellite with a density similar to that of water ice, so the inner satellites migrate inwards as a
result of tidal dissipation within Uranus.

The timescale of
 orbital migration driven by tidal dissipation within Uranus can be
 written as 
 \begin{equation}
 \tau_a=\frac{2}{13}\frac{a}{|\dot a|}=\frac{4}{117}\frac{M_\mathrm{U}}{M_\mathrm{s}}\left(\frac{a}{R_\mathrm{U}}\right)^5\frac{Q'_\mathrm{U}}{n},
 \end{equation}
 where $a$ is the orbital semimajor axis, $n$ is the orbital mean
 motion, $M_\mathrm{U}$ and $R_\mathrm{U}$ are the mass and radius of
 Uranus, $M_\mathrm{s}$ the mass of the satellite and $Q'_\mathrm{U}$
 is the reduced tidal quality factor of Uranus at the relevant tidal
 frequency \citep{Goldreich:1966}. The migration is inward for satellites orbiting inside the corotation radius of the planet and outward for satellites outside corotation. The factor of $2/13$ reflects the accelerating nature of inward migration and gives the time for complete orbital decay under the assumption of constant $Q'_\mathrm{U}$. It is a remarkable fact that
 $\tau_a$ is on the order of $10^5Q'_\mathrm{U}\;\mathrm{yr}$ for
 many of the small inner satellites (including Cordelia and Ophelia)
 that orbit inside the corotation radius as well as for Miranda and Ariel (which are outside corotation), despite the wide range of masses of
 these satellites.  Furthermore, all the regular satellites except
 Cordelia exert tidal forcing on Uranus at a frequency at which
 inertial waves can be excited, assuming that Uranus is uniformly
 rotating \citep{Ogilvie:2004}. Based on evolutionary scenarios for Miranda, Umbriel and Ariel using a frequency-independent tidal quality factor, \citet{Tittemore:1990} have argued\footnote{They state that $11,000<Q_\mathrm{U}<39,000$, and we have $Q'_\mathrm{U}=(3/2)Q_\mathrm{U}/k_2$, where the Love number $k_2=0.104$ for Uranus.} that $Q'_\mathrm{U}$ lies between $1.6\times10^5$ and $5.6\times10^5$, most likely towards the low end of this range.  Although $Q'_\mathrm{U}$ is unlikely to be frequency-independent, values in this range would allow significant inward migration of the smaller inner moons, as well as significant outward migration of Miranda and Ariel, over the age of the solar system.

In order to study tidal disruption and the formation of narrow rings
we have designed a series of numerical simulations. In these
simulations a small satellite,
either homogeneous or differentiated, is placed in a circular orbit in
the gravitational potential of the planet.  The mean density of the
satellite and the orbital semimajor axis are varied. The technical
details of the simulations are described below.

\section{Numerical Method}
The numerical simulations presented in this paper were conducted using
the $N$-body integrator PKDGRAV \citep{Stadel:2001}. PKDGRAV is a
parallelised hierarchical tree-code with a second-order leap-frog
integrator. The satellite is assumed to be a rubble pile -- a
gravitational aggregate bound together by self-gravity
\citep{Richardson:2002}. A rubble pile by definition has no bulk
tensile strength. However, the particles themselves that make up the
rubble pile are rigid spheres and cannot interpenetrate or be crushed.
As a result, the initial rubble-pile satellite has infinite
compressional strength, some shear
strength (due to interlocking particles), and some internal friction
\citep{Leinhardt:2000}.  We have chosen to use a rubble-pile
description for the initial satellites for two reasons: (i)
Cordelia and Ophelia, the satellites on either side of the Uranian
$\epsilon$ ring, are elongated, small bodies whose shape is consistent
with that of a body held together purely by self-gravity \citep[for Cordelia $R =
20$ km, ratio of shortest to longest axes is
0.7,][]{Karkoschka:2001a}; (ii) a rubble pile is the simplest internal
structure we could model in a numerical study of the process of tidal
disruption. The choice of initial satellite mass is largely arbitrary because the
problem is essentially scale-free in the limit of small mass ratio.

The initial satellites are constructed from $4955$ particles
randomly placed in a low-density sphere in isolation (i.e., without a
planetary gravitational potential). 
Self-gravity causes the particles to collapse toward the centre of the
sphere.  In order to damp collisional oscillations, the particle
collisions are made highly inelastic \citep[$\epsilon_\mathrm{n} = 0.1$, see
next paragraph;][]{Leinhardt:2012}. This method of randomisation is
favoured over hexagonal close packing
\citep{Leinhardt:2000,Leinhardt:2002}, which generates a lattice
structure that causes unrealistic results in both tidal disruption and
collisional impacts.

The motion of each particle within the rubble-pile is governed by
gravity and collisions. The inelastic collisions between particles are
parameterised using the normal and tangential coefficients of restitution,
$\epsilon_\mathrm{n}$ and $\epsilon_\mathrm{t}$, respectively \citep{Richardson:1994}.
In these simulations we omit surface friction by setting $\epsilon_\mathrm{t} =
1$, as this would otherwise introduce an additional and poorly
constrained parameter that does not have a significant effect on tidal
disruption events \citep{Richardson:1998}. The normal coefficient of
restitution can vary between 0 (perfectly inelastic) and 1 (perfectly
elastic) and determines the rebound velocity as the result of a
collision. The relative velocity of two particles before their
collision is
\begin{equation}
\mathbf{v} = \mathbf{v}_\mathrm{n} + \mathbf{v}_\mathrm{t},
\end{equation}
where $\mathbf{v}_\mathrm{n}$ is the component of $\mathbf{v}$ normal to the
impact surface and $\mathbf{v}_\mathrm{t}$ is the component tangent to the
impact surface. The relative velocity after the collision is then
\begin{equation}
\mathbf{v'} = -\epsilon_\mathrm{n} \mathbf{v}_\mathrm{n} + \mathbf{v}_\mathrm{t}.
\end{equation}
The energy lost in a collision (to heat or minor deformation of the
rubble pile constituents) is given by
\begin{equation}
\Delta E = -\frac{1}{2}\mu(1-\epsilon_\mathrm{n}^2)v_\mathrm{n}^2,
\end{equation}
where $\mu$ is the reduced mass of the particles involved in the
collision.

Once the rubble piles have been successfully formed we model the inelastic collisions between particles with a higher coefficient of restitution, $\epsilon_\mathrm{n} = 0.5$, which
is consistent with results from laboratory impact experiments using
ice, rock, and soil \citep{Chau:2002,Higa:1996,Higa:1998}. In order to
prevent inelastic collapse within the rubble piles, however, we set
$\epsilon_\mathrm{n} = 1$ for impacts with relative speeds less than 1\% of the
mutual escape speed. As a result the rubble-pile satellites have a low
vibrational temperature.

We have chosen two simplistic internal structure models for the
satellites: (i) a homogeneous body of uniform bulk density; (ii) a
differentiated body consisting of two layers of different bulk
densities.  Disruption of a differentiated body may be able to produce
one or two rings and a remnant satellite, as the denser core may
resist tidal disruption at a semi-major axis where the lower-density
mantle is stripped. The differentiated satellites have a density ratio
of four between the core and mantle and a core size of about $35\%$ by number ($1714/4955$).  Since all the particles have the
same size, the core particles are four times more massive than the
mantle particles. This situation is
more extreme than would be found in nature but has been chosen to make
the results more straightforward to interpret.  For both homogeneous
and differentiated satellites we have investigated six mean densities
from $0.5 - 3.0$ g cm$^{-3}$ and a range of orbital semi-major axes in
order to locate the onset of tidal disruption.

Once the model satellites have been formed they are placed in the
gravitational potential of a central point mass (e.g., Uranus) in a
circular orbit and in synchronous rotation. The system is then integrated for at
least 100 orbits. In general the tidal disruption process is gentle
and the particles within the rubble pile collide with each other
slowly. As a result, we do not expect a large amount of damage or
fracturing to the individual particles. Therefore, the rigid-sphere
model for the constituent rubble-pile particles is a reasonable
approximation to reality.

\section{Results}

We have divided our results into two sections: the first (\S \ref{sec:satdisp}) determines the conditions required for the disruption of a satellite, while the second (\S \ref{sec:ringform}) describes the dynamics of the disruption and the formation of a ring system. We begin our discussion of the satellite disruption by quantifying the disruption criteria for satellites of varying mean densities and internal structure (\S \ref{sec:criteria}). Subsequently, we discuss the various disruption paradigms our simulations exhibit, and connect them to appropriate analytical analogues, namely the evolution of a viscous fluid ellipsoid and the Roche-lobe overflow problem (\S \ref{sec:disdyn}). After the initial disruption phase, a proto-ring system forms, the global structure and evolution of which we relate to the different initial compositions of the disrupted satellite (\S \ref{sec:ringform}).

\begin{figure*}
\includegraphics[scale=0.6]{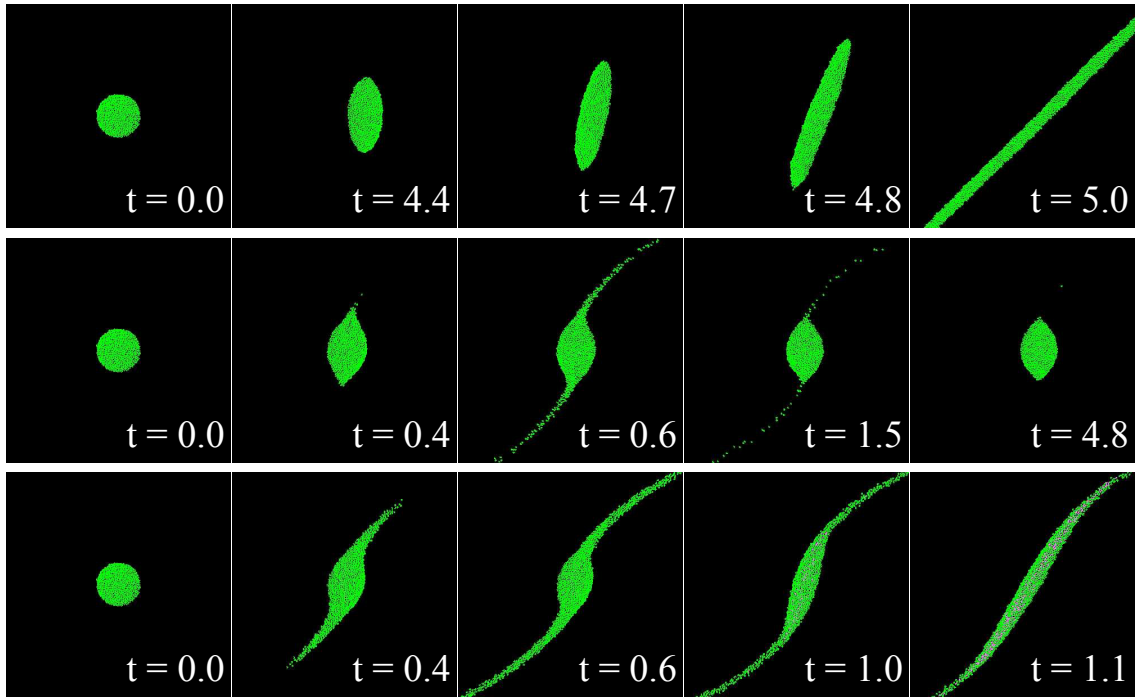}
\caption{Examples of tidal disruption.
 The first row shows the full disruption of a homogeneous satellite
 ($\bar\rho_\mathrm{s} = 0.5$ g cm$^{-3}$, $a = 2.95\, R_\mathrm{p}$).
 The second is an example of mantle disruption for a differentiated
 satellite ($\bar\rho_\mathrm{s} = 0.5$ g cm$^{-3}$, $a = 2.55\,
 R_\mathrm{p}$). The third shows the full disruption of a
 differentiated satellite ($\bar\rho_\mathrm{s} = 0.5$ g cm$^{-3}$,
 $a = 2.36\, R_\mathrm{p}$). In the differentiated satellites the
 mantle particles are shown in green and the core particles are shown
 in magenta. Time increases from left to right and is shown in
 orbital periods. The images are in the rotating frame. 'Up' points
 radially away from the planet and 'down' points radially towards the
 planet and 'left' is along the orbit.}\label{fig:examples}
\end{figure*}

\subsection{Satellite Disruption}\label{sec:satdisp}

\subsubsection{Disruption Criteria}
\label{sec:criteria}

Fig.~\ref{fig:examples} shows examples of the three types of tidal
disruption that occur in the numerical simulations. The top row shows
the full disruption of a homogeneous satellite. The initial change in
shape takes several orbits; particles in the interior cannot move
until particles on the surface have moved out of the way. Once the
satellite has begun to elongate the
disruption accelerates.  The second row
shows an example of a mantle disruption of a differentiated satellite.
In this case only the surface layer particles are moving
significantly.  First the mantle changes into a lemon shape, then
particles flow off the surface through the Lagrange points. This loss
of mantle particles increases the mean density of the satellite as
the core does not change shape significantly nor is there a significant change in the mean density of the core.   Eventually a mean density is reached that is stable
against tidal disruption and mantle particles are no longer lost from
the surface. Note that the mantle stripping process observed here is comparable to that found in \citet{Canup:2010}, implying that the dynamics of the disruption are common across a wide regime of gravity-dominated objects. The last row shows the full disruption of a
differentiated satellite. Initially the mantle begins to disrupt in a
similar but more vigorous fashion to the pure mantle disruption
shown in the row above, with many particles flowing through the
Lagrange points at once.  However, the remnant is not dense enough in
this case and the entire core shears out in a process that is similar
to the disruption of the homogeneous satellite in the top row.

Tables \ref{tab:nocore} and \ref{tab:core} summarise how 
the tidal disruption of homogeneous and differentiated satellites depend on semi-major axis and bulk density.
The result of the simulation is indicated by a D (full disruption; top
and bottom rows of Fig.~\ref{fig:examples}), MD (mantle disruption;
middle row of Fig.~\ref{fig:examples}), or N (no disruption) in the
appropriate square. In cases of mantle disruption or no disruption the
aspect ratio (ratio of shortest to longest axes) of the remnant is
indicated under the disruption type. The axis ratios are determined by
first identifying the particles within the rubble pile using a
clump-finding algorithm \citep{Leinhardt:2000} and then calculating
the moment-of-inertia tensor of the rubble pile
\citep{Richardson:1998}.

\begin{figure*}
\begin{center}
\includegraphics{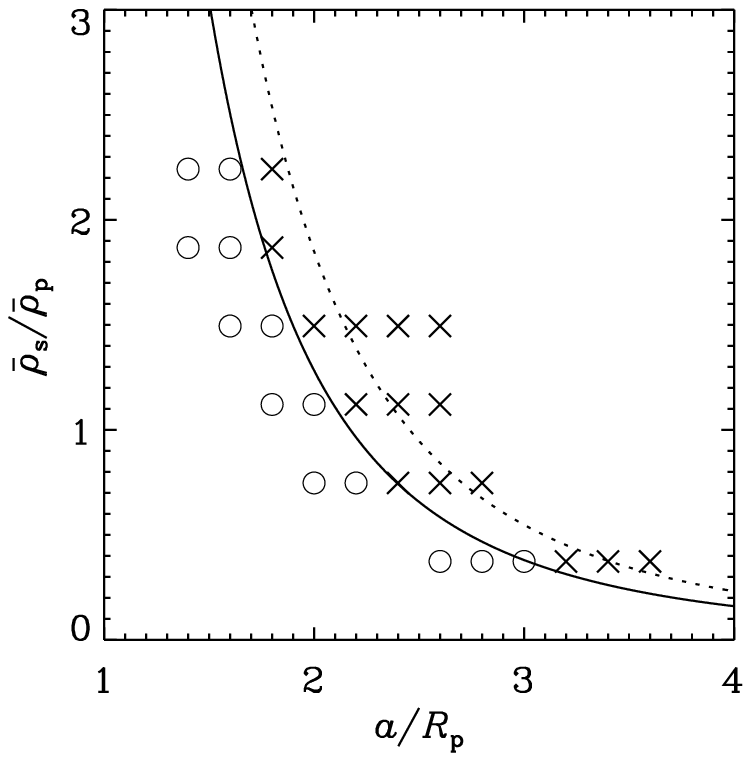}
\includegraphics{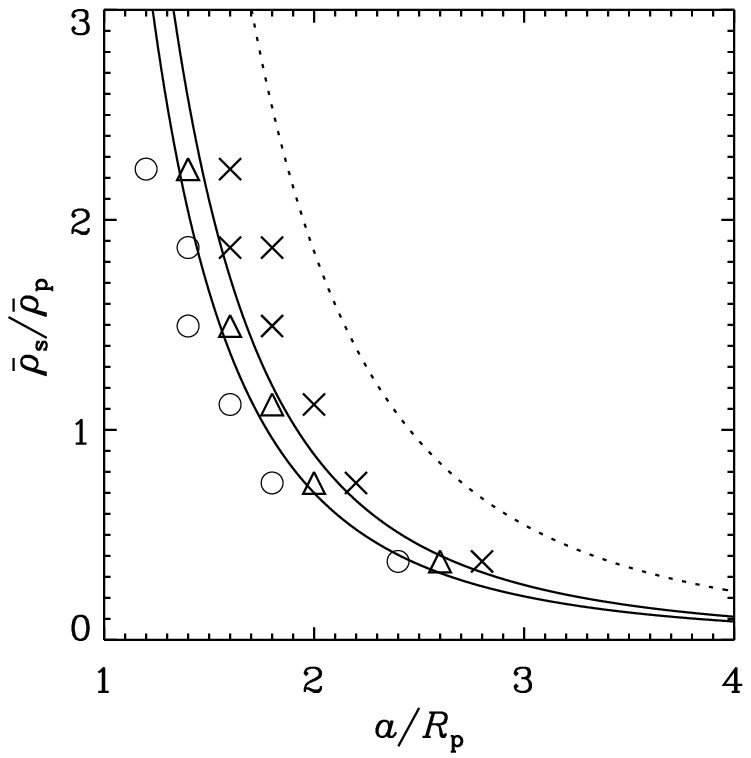}
\end{center}
\includegraphics{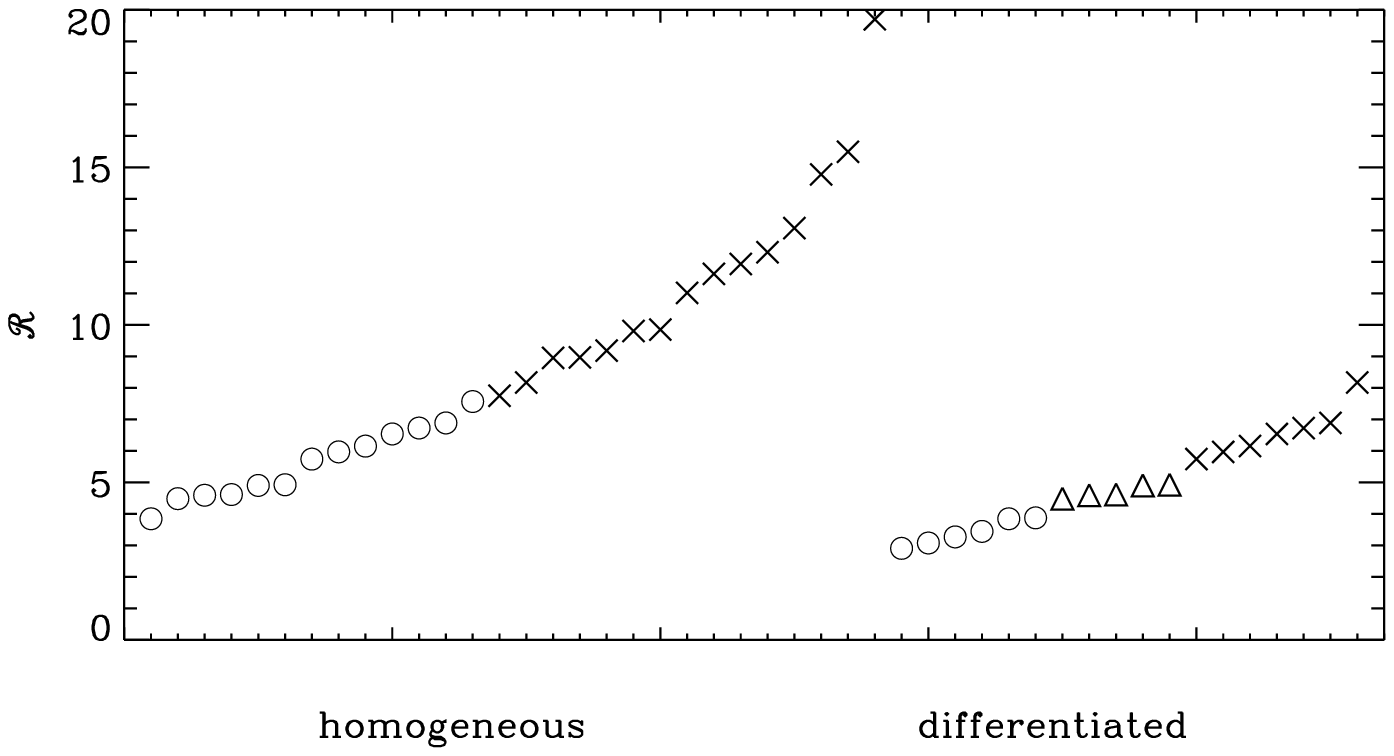}
\caption{Tidal disruption of satellites.  Top left: Outcome of
 numerical simulations of homogeneous satellites for various ratios
 of orbital semi-major axis to planetary radius and various ratios of
 satellite density to planetary density. $\bigcirc$ indicates full tidal
 disruption, while $\times$ indicates no disruption.  The dotted line
 shows the Roche limit for an incompressible fluid body,
 $\mathcal{R}=11.10$, while the solid line shows $\mathcal{R}=7.7$.
 Top right: Similar results for differentiated satellites.  Here
 $\triangle$ indicates mantle disruption only and the two solid
 lines show $\mathcal{R}=4.2$ and $\mathcal{R}=5.3$.  Bottom: Results
 expressed in terms of the Roche parameter.}
\label{fig:disruption}
\end{figure*}

The results are also presented graphically in
Fig.~\ref{fig:disruption}.  The relevant dimensionless parameter in
this problem is the `Roche parameter'
\begin{equation} 
 \mathcal{R}=\frac{\pi G\bar\rho_\mathrm{s}}{\Omega^2}=\frac{3}{4}\frac{\bar\rho_\mathrm{s}}{\bar\rho_\mathrm{p}}\left(\frac{a}{R_\mathrm{p}}\right)^3,
\end{equation}
where $\bar\rho_\mathrm{s}$ and $\bar\rho_\mathrm{p}$ are the mean
densities of the satellite and the planet, respectively, $\Omega$ and
$a$ are the angular velocity and the semi-major axis of the circular
orbit, and $R_\mathrm{p}$ is the volumetric radius of the planet (25362 km for Uranus).
For an incompressible fluid, tidal disruption occurs for
$\mathcal{R}<11.10$, where a solution in the form of a Roche
ellipsoid can no longer be found \citep[e.g.][]{Chandrasekhar:1969};
this condition corresponds to
\begin{equation}\label{eqn:roche}
 \frac{a}{R_\mathrm{p}}<2.46\left(\frac{\bar\rho_\mathrm{s}}{\bar\rho_\mathrm{p}}\right)^{-1/3}.
\end{equation}
%In the case of Uranus,
%$\bar\rho_\mathrm{p}=1.27\,\mathrm{g}\,\mathrm{cm}^{-3}$.

It is to be expected that rubble piles can survive to smaller values
of $\mathcal{R}$ than incompressible fluid bodies because of their
shear strength.  Indeed, for the homogeneous satellites we find a
critical disruption distance that is closer to the planet than
described by equation~(\ref{eqn:roche}).  We find a coefficient of
about 2.2 instead of 2.46 (corresponding to a critical Roche parameter
of about 7.7, as shown in Fig.~\ref{fig:disruption}).  This coefficient is still larger
than would be expected from the work of \citet{Holsapple:2006} and
\citet{Holsapple:2008}, which predicts a disruption coefficient of 1.7
for a homogeneous elongated body with an axis ratio of $\alpha = 0.7$,
no tensile strength, and a friction angle of $\phi = 20^\circ$. We
must be careful here because the criteria presented in
\citet{Holsapple:2006} do not distinguish between a change in shape
and true disruption; their calculations give only necessary conditions
for disruption and not sufficient conditions. Therefore, perhaps the
more reasonable comparison with \citet{Holsapple:2006} is to consider
the critical distance within which the satellite necessarily departs
from a sphere. We find that a rubble-pile satellite can maintain a
spherical shape at a distance $a/R_\mathrm{p} = (2.80\pm0.15)\,
(\bar\rho_\mathrm{s}/\bar\rho_\mathrm{p})^{-1/3}$. Within this
limiting radius the rubble pile reshapes due to the tidal forces and
transforms from a sphere into an ellipsoid. Thus, our rubble-pile
satellites behave somewhere in between a fluid and a rigid solid. Note that
if the satellites were true fluids there would be \emph{no} distance
away from the central potential where a true spherical shape could be
maintained. In this respect we are in agreement with
\citet{Holsapple:2006} who also predict the existence of spherical
equilibria outside a critical orbital radius.  They find, however,
that this radius is between $1.8$ and $ 2.2$ times $(\bar\rho_\mathrm{s}/\bar\rho_\mathrm{p})^{-1/3}$ (for friction
angles between 10 and 20 degrees). So there is a
discrepancy of roughly 20--35\%, which is acceptable given the
uncertainties in their failure criterion and in its strict
applicability to our numerical model.

The lower critical Roche parameter for the differentiated satellite
can be explained, at least in part, by its greater central
condensation.  In the limit of a completely centrally condensed fluid
satellite, the critical Roche parameter is approximately 6.25, which is significantly lower than the value of 11.10 for a homogeneous fluid body.  The value 6.25 follows from the fact that the volume of the Hill sphere is approximately 1.509 Hill units (as determined by numerical integration).

\begin{table*}
\caption{Tidal disruption table for homogeneous satellites.\label{tab:nocore}}
\begin{tabular}{|*{13}{c|}}
\hline
$\bar \rho$ & 1.37 $R_\mathrm{p}$ & 1.57 $R_\mathrm{p}$ & 1.77 $R_\mathrm{p}$ & 1.96
$R_\mathrm{p}$ & 2.16 $R_\mathrm{p}$ & 2.36 $R_\mathrm{p}$ & 2.55 $R_\mathrm{p}$ & 2.75 $R_\mathrm{p}$ & 2.95 $R_\mathrm{p}$ & 3.14 $R_\mathrm{p}$ & 3.34 $R_\mathrm{p}$ & 3.54 $R_\mathrm{p}$ \\
\hline
0.5 & & & & & & & \multirow{2}{*}{D} & \multirow{2}{*}{D} & \multirow{2}{*}{D} & \multirow{2}{*}{N} & \multirow{2}{*}{N} & \multirow{2}{*}{N}\\
 & & & & & & &  & & & 0.74 & 0.82 & 0.88\\
\hline
1.0 & &  & & \multirow{2}{*}{D} & \multirow{2}{*}{D} & \multirow{2}{*}{N} &  \multirow{2}{*}{N} &  \multirow{2}{*}{N} &&&&\\
 & & & & & & 0.55 & 0.77 & 0.86 &&  & &\\
\hline
1.5 & & & \multirow{2}{*}{D} & \multirow{2}{*}{D} & \multirow{2}{*}{N} &  \multirow{2}{*}{N} &  \multirow{2}{*}{N} & &&&&\\
 & & & & & 0.7 & 0.82 & 0.91 & &&&&\\
\hline
2.0 &  & \multirow{2}{*}{D} & \multirow{2}{*}{D} & \multirow{2}{*}{N} & \multirow{2}{*}{N} & \multirow{2}{*}{N} & \multirow{2}{*}{N} & &&&&\\
 & & & & 0.71 & 0.84 & 0.90 & 0.92 & &&&&\\
\hline
2.5 &\multirow{2}{*}{D} & \multirow{2}{*}{D} & \multirow{2}{*}{N} & & & & & &&&&\\
 & & & 0.67 & & & & & &&&&\\
\hline
3.0 & \multirow{2}{*}{D} & \multirow{2}{*}{D} & \multirow{2}{*}{N} & & & & & &&&&\\
 & & & 0.81& & & & & &&&&\\
\hline
\end{tabular}
\end{table*}

\begin{table*}
\caption{Tidal disruption table for differentiated satellites.\label{tab:core}}
\begin{tabular}{|*{11}{c|}}
\hline
$\bar \rho$ & 1.18 $R_\mathrm{p}$ & 1.37 $R_\mathrm{p}$ & 1.57 $R_\mathrm{p}$ & 1.77 $R_\mathrm{p}$ & 1.96
$R_\mathrm{p}$ & 2.16 $R_\mathrm{p}$ & 2.36 $R_\mathrm{p}$ & 2.55 $R_\mathrm{p}$ & 2.75 $R_\mathrm{p}$\\
\hline
0.5 & & & & & & & \multirow{2}{*}{D} & \multirow{2}{*}{MD}&\multirow{2}{*}{N} \\
 & & & & & & &  & 0.56 & 0.64\\
\hline
1.0 & &  & & \multirow{2}{*}{D} & \multirow{2}{*}{MD} & \multirow{2}{*}{N} & & &\\
 & & & & & 0.53 & 0.64 & & &\\
\hline
1.5 & & &\multirow{2}{*}{D} & \multirow{2}{*}{MD} & \multirow{2}{*}{N} & & & &\\
 & & & & 0.56 & 0.74 & & & &\\
\hline
2.0 &  &  \multirow{2}{*}{D} &\multirow{2}{*}{MD} & \multirow{2}{*}{N} & & && & \\
 & & & 0.59 & 0.73 & & & & &\\
\hline
2.5 & &\multirow{2}{*}{D} & \multirow{2}{*}{N} & \multirow{2}{*}{N} & & & & & \\
 & & & 0.54 & & & & & &\\
\hline
3.0 & \multirow{2}{*}{D} & \multirow{2}{*}{MD}& \multirow{2}{*}{N}& & & & & & \\
 & & 0.58 & 0.76 & & & & & &\\
\hline
\end{tabular}
\end{table*}

\subsubsection{Disruption Dynamics}\label{sec:disdyn}

Turning now to the dynamics of tidal disruption, we compare the
outcomes of our $N$-body simulations of tidal disruption with two
simplified analytical models.  The disruption of a homogeneous
satellite can be compared with the behaviour of a homogeneous viscous
fluid ellipsoid which is described in detail below.  On the other hand, the disruption of the
mantle from an inhomogeneous satellite can be compared with Roche-lobe
overflow which is described later in this section.

{\it Homogeneous Satellite: Disruption of a Viscous Fluid Body}\label{section:nocore}

\begin{figure*}
\includegraphics[scale=1.0]{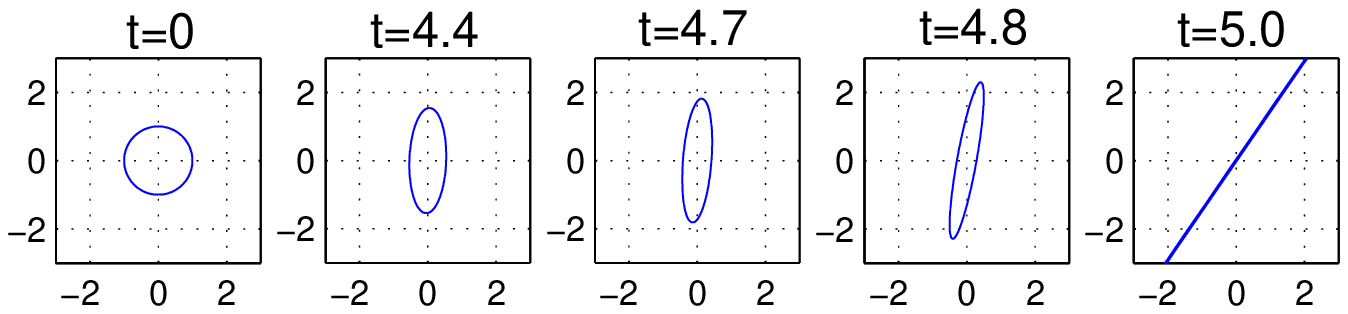}
\caption{Four snapshots of the evolution of a viscous fluid body
 within its Roche limit, as computed by the ST92 model in a circular
 orbit. The Roche parameter is $\mathcal{R}= \pi G \rho/\Omega^2 =
 7.57$ and the alpha viscosity parameter is $\alpha=5.0$. The
 times are in orbital periods but the unit of length is arbitrary.
 `Down' is towards the planet, and `left' is along the circular
 orbit. The initial state is a sphere. This figure should be compared
 with the first row of Fig.~\ref{fig:examples}.}\label{fig:ellipsoid}
\end{figure*}

In this section we employ a simple fluid model that helps to
illuminate the salient physics of satellite disruption. As
\citet[][hereafter ST92]{Sridhar:1992} argue, a homogeneous rubble
pile, exhibiting compressive and shear strengths but no tensile
strength, can be crudely approximated by an incompressible viscous
fluid ellipsoid.  Under the influence of tidal forces, self-gravity,
pressure, and viscous stresses, the shape and orientation of this
ellipsoid can be computed with minimal effort, and its ultimate
disruption reproduced.  It is thus a convenient tool with which to
interpret the more involved $N$-body simulations.

The formalism we use is that of ST92, and it consists of two sets of
ordinary differential equations: the first, derived from the
Navier--Stokes equations, describes the time-evolution of the
constrained motions within the body; the second, derived from the
boundary conditions, describes the evolution of the ellipsoidal
geometry. Because we add little to the original formalism, these
equations are not derived or listed: the reader is referred to their
elegant presentation in ST92. The main difference is that our
satellite is in a circular orbit at a fixed radius, not the parabolic
orbit that ST92 examine. Consequently, the equations are solved in a
corotating Cartesian frame centred upon the satellite that includes
the Coriolis and centrifugal forces. We denote by $\Omega$ the orbital
frequency at this radius, and the satellite is assumed to be initially
spin-locked.

To approximate the internal friction of the rubble pile a viscous
stress is introduced with the dynamic viscosity proportional to the
pressure. This mimics the expected increase in the shear stresses due
to an applied pressure that is characteristic of most geological and
granular materials \citep{Holsapple:2006, Holsapple:2008}. Simply put,
the more the rubble pile is `squeezed', the greater its resistance to
shear. Thus we have $\eta = \alpha(P/\Omega)$, where $\eta$ is the
dynamic viscosity, $P$ is pressure, and $\alpha$ is a dimensionless
parameter. There is a second parameter, the `Roche parameter'
$\mathcal{R}= \pi G \rho/\Omega^2$ introduced in
Section~\ref{sec:criteria}, where $\rho$ is the (constant) mass
density of the satellite.

In order to compare with the $N$-body simulations, we set the distance
between the satellite and Uranus to be $3R_\mathrm{p}$ and take $\rho=
0.5$ g cm$^{-3}$. This gives $\mathcal{R}=7.57$ (well within the
Roche limit). On the other hand, the viscosity parameter $\alpha$ is
difficult to constrain, and we tried various values. But a higher
$\alpha = 5$ yielded the best comparison. The initial state
was usually a sphere.

In all our runs the evolution followed the same template: (a) a slow
initial elongation of the body towards a roughly 2:1:1 configuration
pointed at the planet, and then (b) a rapid extension in the azimuthal
direction. Eventually the body transforms into a `needle' with one of
the principal axes of the ellipsoid orders of magnitude greater than
the other two. This dominant axis is ultimately aligned with the
azimuthal direction. Accompanying this change are nonlinear
oscillations in the internal motions, which may seed the growing
perturbations observed in the $N$-body simulations (see \S
\ref{sec:ringform} or Fig.~\ref{fig:rings} below). The viscosity
parameter $\alpha$ strongly influences the slow first stage of the
evolution: a large value can prolong this stage significantly.
However, once the second shearing phase begins, the viscosity and
pressure become unimportant and the fluid behaves more like a
collection of ballistic particles.

In Fig.~\ref{fig:ellipsoid}, we plot some snapshots of a typical
evolution, to be compared with the first row of Fig.~\ref{fig:examples} describing the $N$-body
simulations. The agreement between the two approaches is excellent and
emphasises some interesting points. The peculiarities of the granular
flow are unimportant in both stages of the evolution: in the first
stage, all that matters is that there is some kind of internal
friction, and its details are not essential; in the second stage, the
particles behave ballistically and their collective dynamics is
governed by the tidal force and the waning self-gravity force.
Collisions are subdominant. What is striking, perhaps, is how far into
the disruption the $N$-body simulation behaves like the incompressible
fluid. The body does not break up into a `gaseous' belt of relatively
energetic particles -- that comes later -- but remains in a coherent
closely packed `fluid' phase for a substantial fraction of the ring
formation. As is shown in \S \ref{sec:ringform} below, instabilities
finally destroy this phase.

{\it Differentiated Satellite: Roche-Lobe
 Overflow}\label{sec:differentiated}

\begin{figure*}
\includegraphics[scale=0.6]{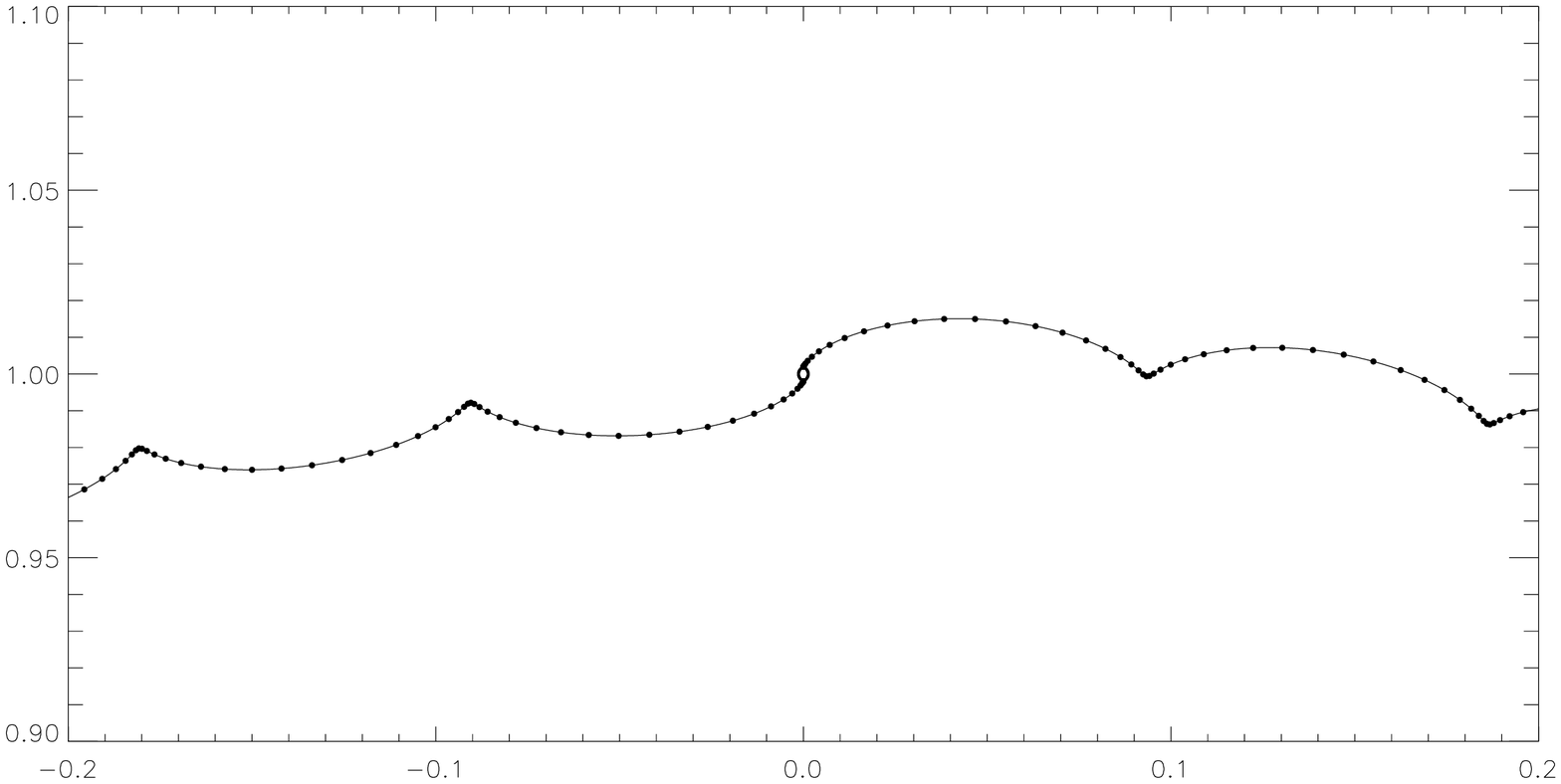}
\caption{Trajectories of particles in the circular restricted
 three-body problem that start at rest close to the Lagrangian points
 L1 and L2.  Twenty particle positions per orbital period are
 superimposed on the trajectories, and the Roche lobe is also shown.
 The mass ratio is $2.335\times10^{-8}$.  Coordinates are referred to
 the central mass and are in units of the orbital separation.}
\label{fig:threebody}
\end{figure*}

\begin{figure*}
\includegraphics[scale=0.6]{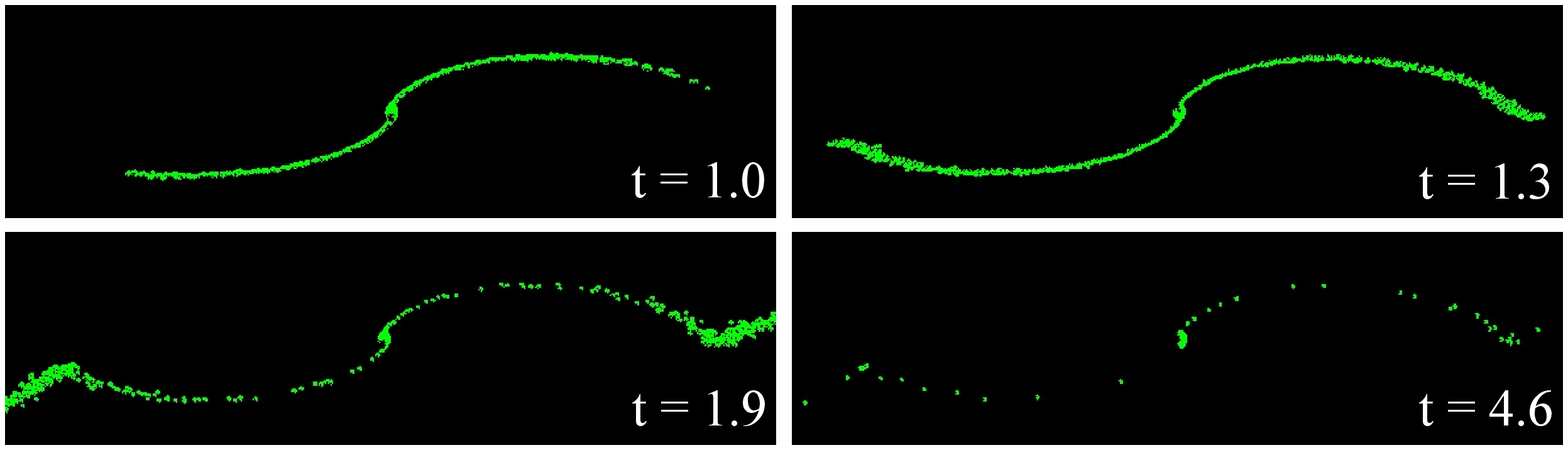}
\caption{Zoomed-out view of the same simulation shown in the second
 row of Fig.~\ref{fig:examples}. The frame is $\sim 0.3$ units of the orbital separation in length, $\sim 340$ Hill radii of the original satellite (1 Hill radius $\sim$ 70 km).  
 The sizes of particles have been inflated by a factor of ten in order to make
 them more visible. This figure should be compared with
 Fig.~\ref{fig:threebody}.}
\label{fig:mantlenum}
\end{figure*}

Fig.~\ref{fig:examples} shows clearly that a differentiated satellite
with a dense core behaves differently from a homogeneous satellite. In
the case of the mantle disruption (middle row) the satellite becomes
elongated into a lemon shape and then begins to lose mantle particles
through the Lagrange points L1 and L2. The mass loss slows when the
mean density of the satellite is high enough for the satellite to
resist disruption \citep[see also][]{Canup:2010}.

The motion of the ejected material can be modelled by considering the
circular restricted three-body problem.  In these calculations we use
the same mass ratio as in the $N$-body simulations.  Particles are
started at rest close to the Lagrangian points L1 and L2 so that they
accelerate away from the Roche lobe (or Hill sphere) and approach Keplerian
orbits far from the satellite.  The trajectories are plotted in the
frame that rotates with the two-body orbit in
Fig.~\ref{fig:threebody}.  In this frame the particles nearly come to
rest once per orbital period.  Eventually the particles encounter the
satellite again.

This calculation can also be done within the Hill approximation, in
which case the final state of the ejected particles (not allowing for
any subsequent encounter with the satellite) is an epicyclic motion with
an amplitude of $3.08$ Hill radii and a guiding centre displaced by
$4.96$ Hill radii from the orbit of the satellite.  The motion
therefore extends from $1.88$ to $8.04$ Hill radii.

A zoomed-out view of the $N$-body simulation of mantle disruption is
shown in Fig.~\ref{fig:mantlenum}.  In a frame rotating with the
satellite's orbit, the streams of ejected particles form rounded
scallop shapes as they trace out the trajectories of the three-body
problem described above.  The density of the tidal streams is enhanced
at the locations where the particles nearly come to rest and
trajectories turn around.  Here the particles collide with one another
and their individual
trajectories are changed (see second and third frame of
Fig.~\ref{fig:mantlenum}). By the last frame in
Fig.~\ref{fig:mantlenum} the mass loss from the mantle has reduced
significantly. Particles collide much less frequently and the scallop
shape becomes cleaner.

As shown in the last row of Fig.~\ref{fig:examples}, a differentiated
satellite that is close enough to the planet will be fully disrupted.
This process shows characteristics of both the disruption of the
homogeneous satellite and the mantle disruption discussed above. At
first the mantle material flows off of the satellite at locations
close to the Lagrange points but the stream created by the mantle
particles is much thicker than in the mantle-only disruption event.
Once most of the mantle is removed, the core also disrupts and
stretches out into an elongated needle similar to the disruption of
the homogeneous satellite.

\subsection{Ring Formation and Ring Properties}\label{sec:ringform}

\begin{figure*}
\includegraphics[scale=0.6]{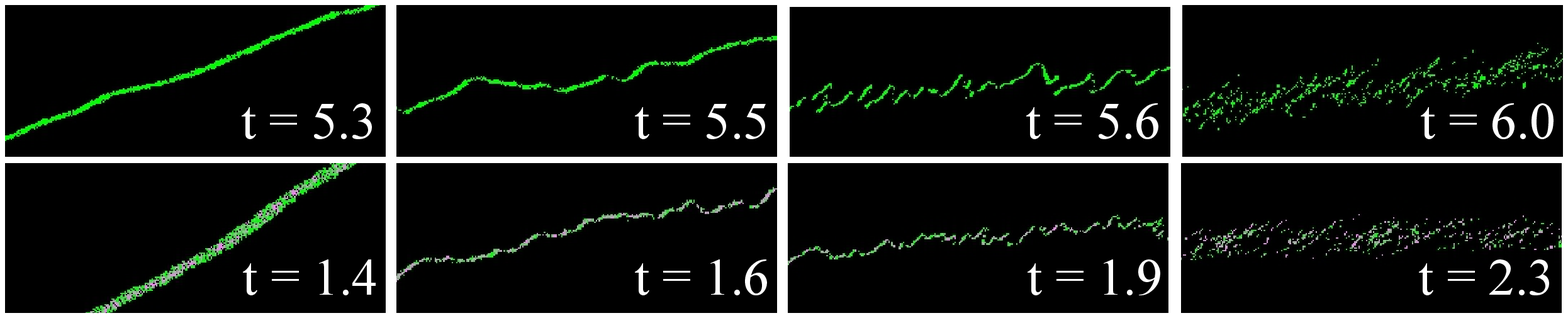}
\caption{Snapshots of the formation of rings from two fully disruptive
 simulations. These frames
 follow on from the first and third rows of Fig.~\ref{fig:examples}:
 the top row is for a homogeneous satellite and the bottom row is for
 a differentiated satellite in which the denser core particles are
 coloured magenta. In the top row the frames from left to right have lengths of 
 13, 26, 53 and 53 Hill radii of the original homogeneous satellite. In the bottom row the frames have lengths of 16, 33, 66, and 66 Hill radii of the original differentiated satellite. The size of the frames was chosen to have the same physical size (1000, 2000, 4000, and 4000 km).}\label{fig:rings}
\end{figure*}

As remarked in Section \ref{section:nocore}, the initial disruption of
the homogeneous satellite is remarkably ordered, consisting of the
rapid shearing-out of the body into a coherent, needle-like ellipsoid.
However, after some time, this ellipsoid (or proto-ringlet) is
attacked by a secondary gravitational instability, the evolution of
which is described in Fig.~\ref{fig:rings}. The complete disruption of
the differentiated satellite follows a similar evolution. The
instability kinks the ringlet into a jagged ribbon on small scales,
eventually destroying its coherence through physical collisions or
gravitational encounters.  Gravitational potential energy is injected
into random motions and the ringlet spreads, becoming a `hot' belt of
particles.  Because of the distinct kink-like morphology of the
unstable mode, it is probably not of the same class as the
longitudinal instability that attacks a self-gravitating cylinder
\citep[described by][]{Chandrasekhar:1961}.  It may be, however,
related to the non-axisymmetric instabilities of thin incompressible
rings and tori calculated by various authors \citep[see][]{Cook:1964,
 Goodman:1988, Papaloizou:1989, Christodoulou:1992, Latter:2012}. What is
interesting is that the deterioration of the homogeneous satellite into
a disordered and collisional belt of particles is a two-stage process:
first an ordered shearing-out into a coherent ellipsoidal-like body of
`cold' particles, and then the disruption of this body by a secondary
instability that rips up the ordered object into an incoherent
collisional belt.

\begin{figure*}
\includegraphics[scale=0.8]{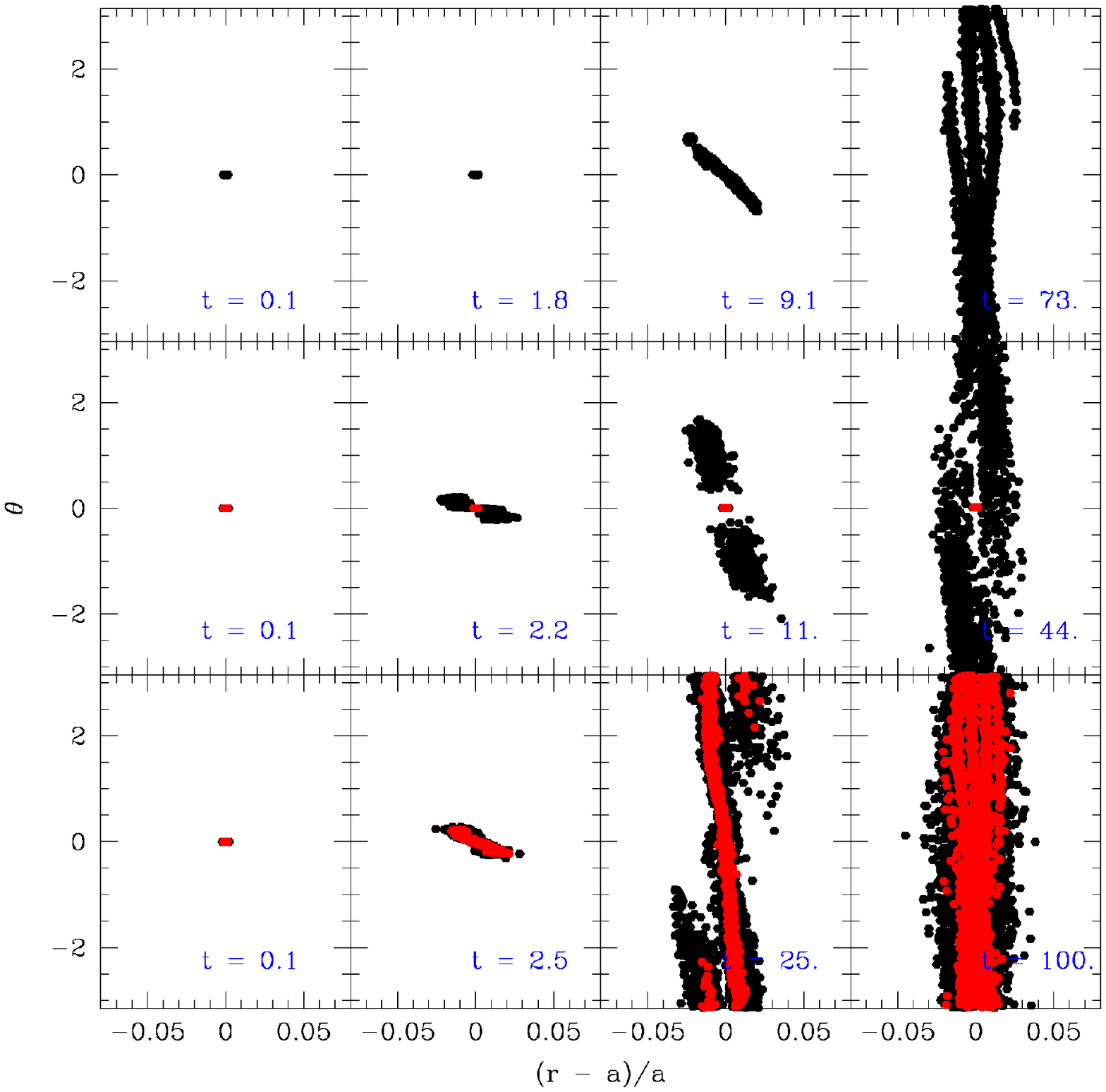}
\caption{Ring particle locations in polar coordinates rotating with
 the satellite's orbit. The core particles are shown in red. The $x$
 axis shows the radial positions of the ring particles with respect
 to the initial location of the satellite, $a$. 
 The three rows correspond to the three simulations shown in
 Fig.~\ref{fig:examples}: the top row is the full disruption of a
 homogeneous satellite, the middle row is the mantle disruption of a
 differentiated satellite, and the bottom row is the full disruption
 of a differentiated satellite.}\label{fig:rotrings}
\end{figure*}

In order to get a more global picture of the longer-term ring
evolution let us look at the ring in polar coordinates rotating with
the satellite's orbit (Fig.~\ref{fig:rotrings}). Note that the
evolution shown in Fig.~\ref{fig:rings} occurs while the disrupted
satellite is still an incomplete ring. In the case of the homogeneous
satellite (top row), the ring arc continues to grow in azimuth until
it wraps around, eventually forming a spiral structure. There is a
periodic sinusoidal pattern due to the eccentricity of the particle
streams, which results from the original location of the particles in the satellite. However, in the case of the mantle disruption of the differentiated
satellite (middle row), the periodic stream structure is destroyed by
the remnant core. The mantle particles are scattered by the core at
every encounter. In the full disruption of the differentiated
satellite (bottom row), the core particles show the same wrapping
periodic pattern of the homogeneous satellite. The mantle particles,
on the other hand, show much less of a coherent pattern because the
core did not disrupt immediately. Thus, some of the mantle particles
were scattered by a still coherent core.

\begin{figure*}
\includegraphics[scale=0.7]{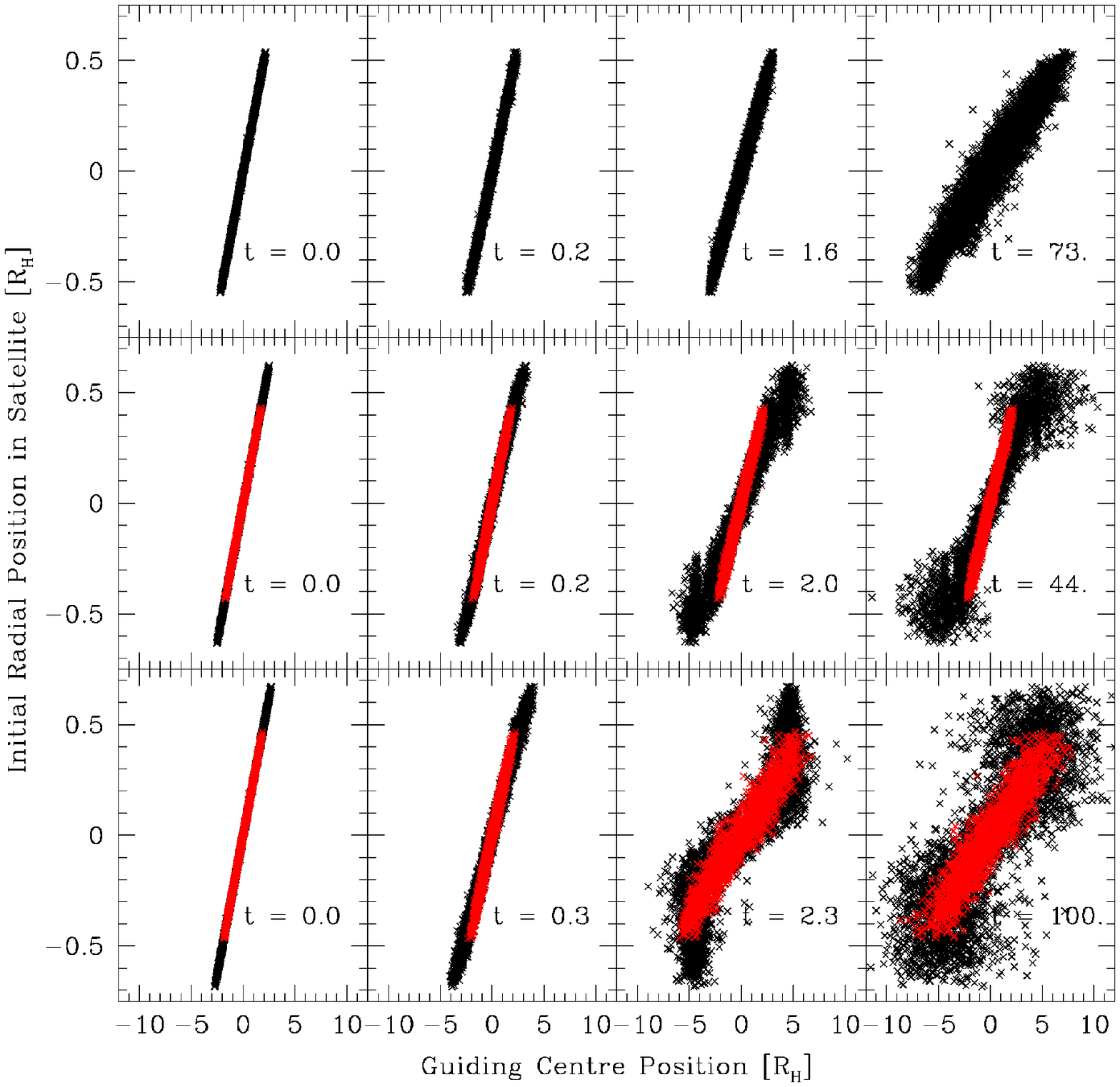}
\caption{The original location of particles in the initial satellite
 versus the guiding centre position of the particles in the ring in units of the Hill radii of the original satellite. The
 core particles are indicated in red. The three rows correspond to
 the three different types of disruption shown in
 Fig.\ref{fig:examples}.  }\label{fig:where}
\end{figure*}

\begin{figure*}
\includegraphics[scale=0.7]{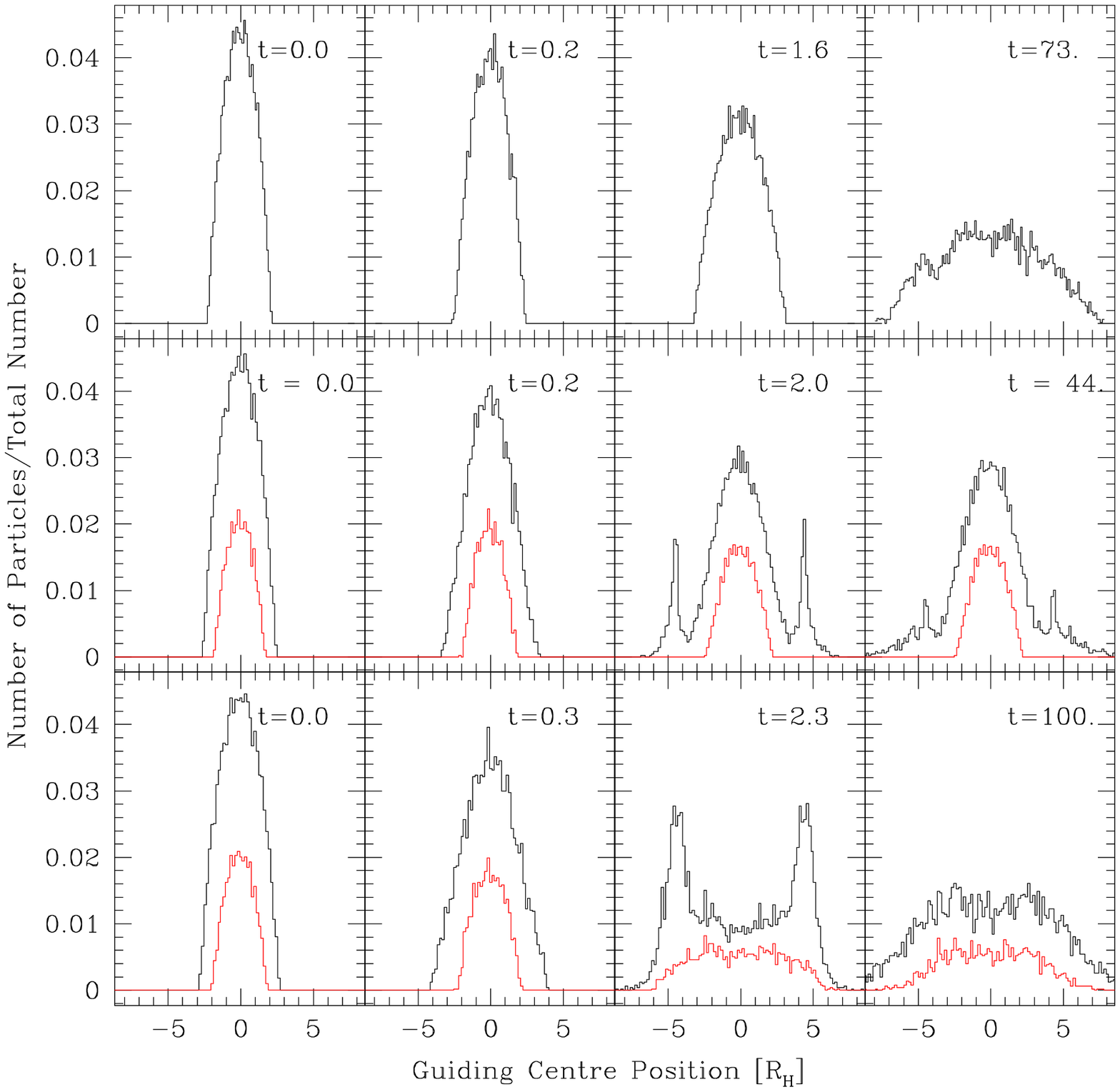}
\caption{Histograms of the guiding centre position. The black line indicates all particles, while the red line is for core particles only.}\label{fig:surface}
\end{figure*}

We now discuss the dynamical state of the newly formed rings.  Each
particle can be considered to execute an elliptical Keplerian orbit,
interrupted infrequently by collisions or gravitational encounters
with other ring particles or the remnant satellite.  The orbital
properties of greatest interest are the specific angular momentum and
the eccentricity.

The specific angular momentum of a ring particle can be expressed in
terms of the radius of the guiding centre of its epicyclic motion,
$r_g = (L_z/m)^2 (G M_\mathrm{p})^{-1}$, where $L_z$ is the $z$
component of angular momentum and $m$ is the mass of the particle.
Fig.~\ref{fig:where} shows the guiding-centre radius of the
particles as a function of the initial radial location within the
satellite.  We have subtracted from each of these the radius of the
initial satellite orbit so that the disruption of satellites initially
placed at different radial locations can easily be compared with each
other on the same axes. The change in global slope of the mantle particles is due to the initial re-equilibration of the rubble pile before disruption. Note in the middle row the core maintains its original slope which is an indication that the core has stayed spherical and has not re-equilibrated. This is also initially the case in the full disruption of the differentiated satellite. Note the change in slope of the core particles lags behind the change in slope of the mantle particles. However, by the second orbit the core has deformed and subsequently fully disrupted. 
The diffusion of the
initial line of particles to a ``fatter'' distribution is the
result of the inter-particle scattering.  This diffusion can also be
seen in the top row of Fig.~\ref{fig:surface}, which shows the number
of particles as a function of guiding centre position. Over time the
distribution of particles broadens and the peak particle number at the
centre of the ring drops smoothly.

In the mantle disruption of the differentiated satellite the particles are
released from L1 and L2 but they then interact gravitationally with
the remaining core creating the lopsided bow-tie structure seen in
last frame of the middle row of Fig.~\ref{fig:where}. Initially the
mantle disruption shows up as a double-horned feature in
Fig.~\ref{fig:surface} but as the particles interact with the core the
peaks broaden into wings in the overall number density distribution
(last frame).

In the full disruption of the differentiated satellite the evolution
of the particles is a bit confused as some of the mantle particles
seem to be released from L1 and L2 but the core and remaining mantle
material follow a disruption path that is similar to the homogeneous
satellite shown in the top row. For example, the third frame of the
last row in Fig. \ref{fig:surface} looks like a combination of the
mantle disruption and the total disruption described above, with the
double-peaked structure of the mantle particles and the broad
structure of the disrupted core. However, this horned feature
disappears and ring spreads. By 100 orbits the distribution of
particles in the ring looks like the homogeneous disruption
shown in the top row. In both the mantle disruption and the total
disruption of the differentiated satellite we can see evidence of ring
spreading and inter-particle scattering.

\begin{figure*}
\includegraphics[scale=0.7,angle=270]{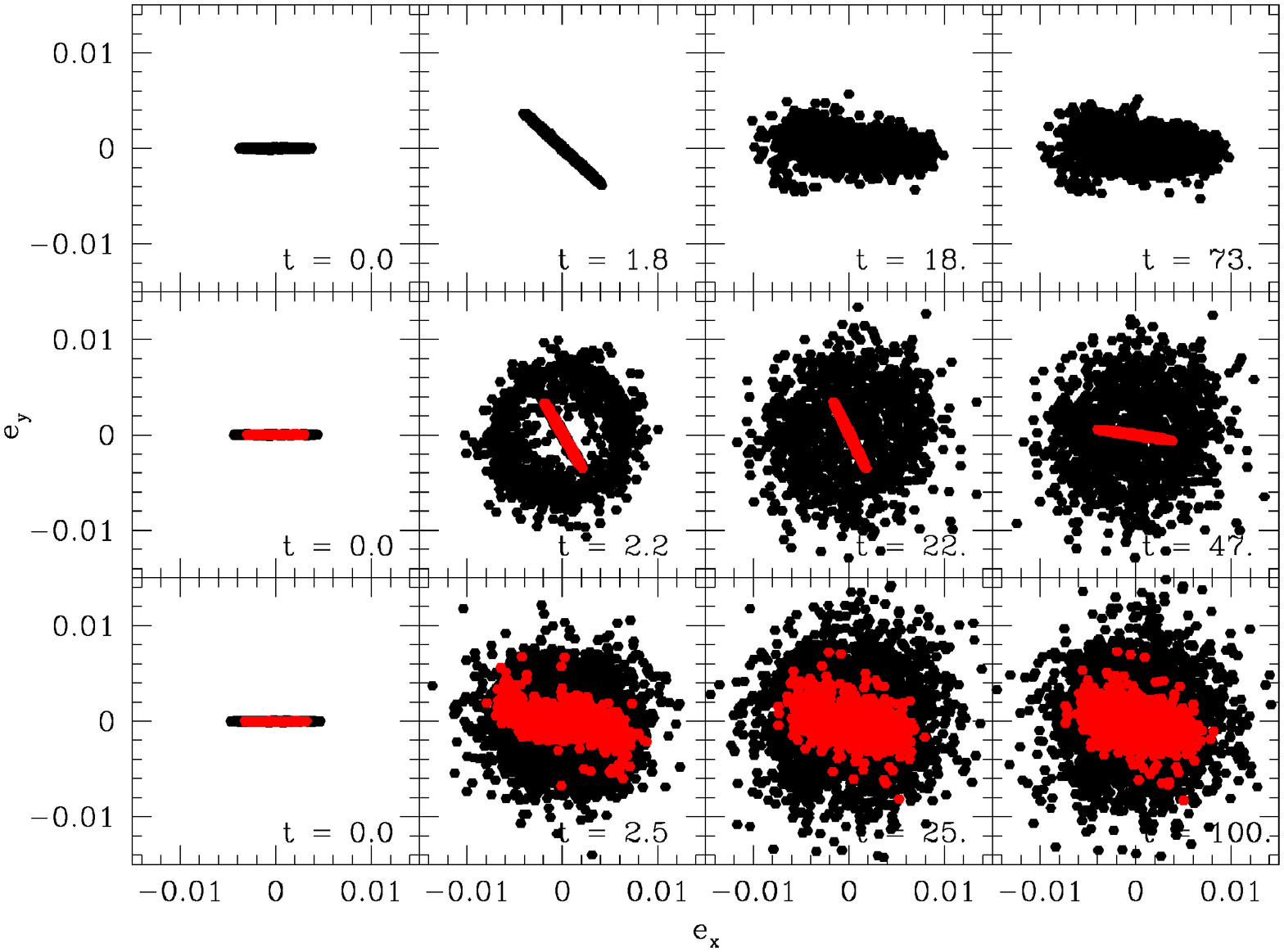}
\caption{Evolution of the eccentricity vectors for all particles versus time. The core particles are indicated in red.}\label{fig:eccvec}
\end{figure*}

\begin{figure*}
\includegraphics[scale=0.7]{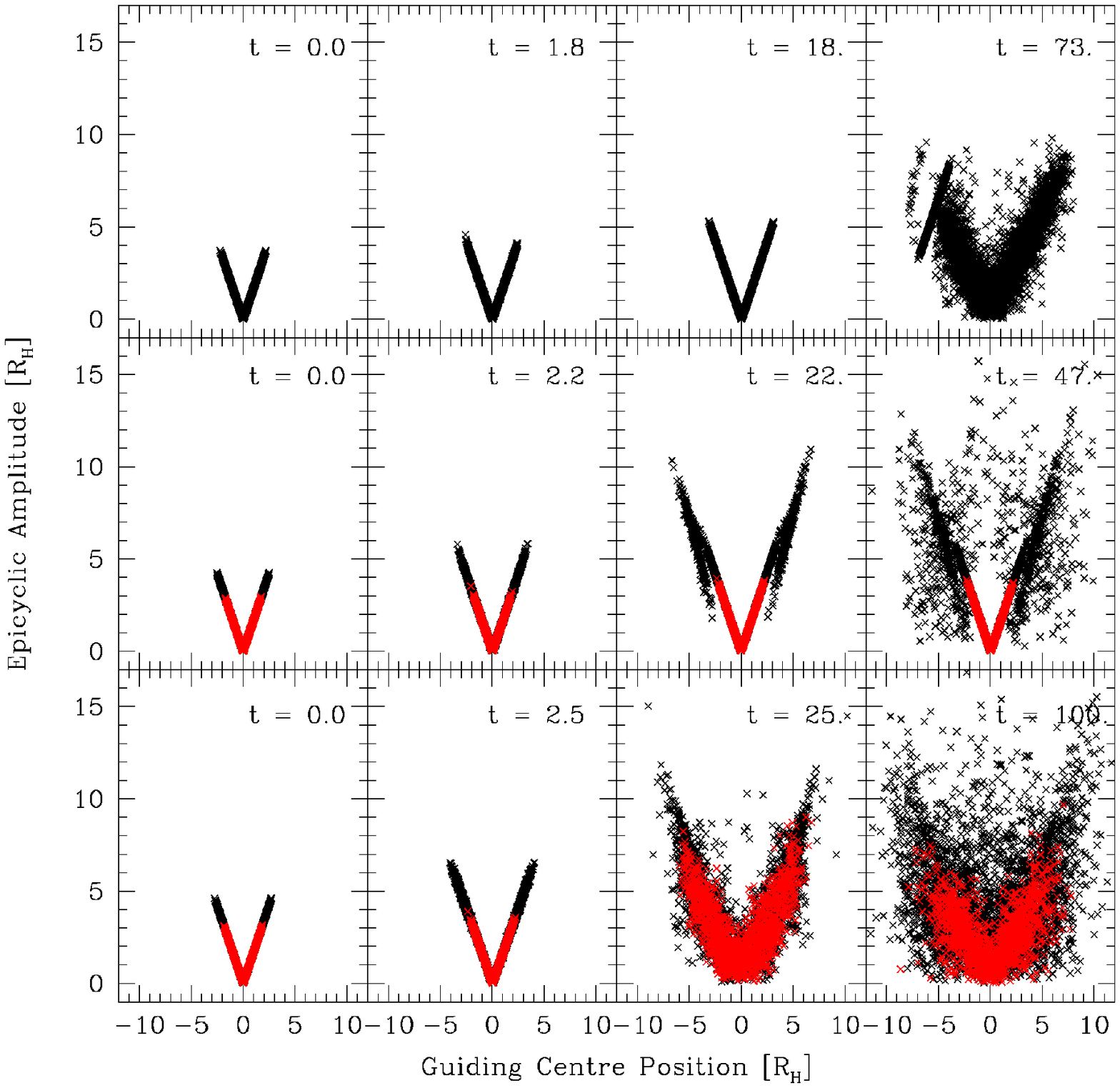}
\caption{Epicyclic amplitude versus guiding centre position. Red indicates core particles.}\label{fig:epiamp}
\end{figure*}

The eccentricity vector and epicyclic amplitude, shown in
Fig.~\ref{fig:eccvec} and Fig.~\ref{fig:epiamp}, respectively, show
the initial dynamical state of the newly formed ring in each type of
disruption event and reveal the extent of the interactions between
ring particles. The homogeneous satellite produces a distribution in
Fig.~\ref{fig:eccvec} with the total extent of $e_y$ equal to about half the total extent of $e_x$, where
$e_x = e\,\cos\varpi$ and $e_y = e\,\sin\varpi$ ($\varpi$ is the longitude of pericentre), and
preserves the characteristic ``v'' shape in the epicyclic amplitude, suggesting that the particles have not interacted with
each other after the disruption. Interestingly the linear anomaly seen
in the last frame in the top row is due to an undisrupted rubble-pile chunk that survived the tidal disruption. In this particular simulation the initial satellite was closer to the critical Roche limit than the other simulations allowing a chuck of material to remain gravitationally intact.

The mantle-only disruption produces eccentricity components that have
the same range (in the $x$ and $y$ components of the eccentricity vector in
Fig.~\ref{fig:eccvec}) suggesting interaction and angular momentum
exchange between mantle particles. In addition, the epicyclic
amplitude shows many more particles outside of the characteristic ``v''
indicating significant interaction with the remnant core.

Finally, the full disruption of the differentiated satellite shows a
combination of the two results. The core material looks vaguely like
the homogeneous satellite in Fig.~\ref{fig:eccvec} and Fig.~\ref{fig:epiamp}
with a rough two-to-one relationship between $e_y$ and
$e_x$ and a 'v' shape in epicyclic amplitude and guiding centre position. The mantle material, on the other hand, is uniformly distributed in eccentricity between
-0.01 and 0.01 and shows a broad range of epicyclic
amplitude and guiding centre position. This is a result of mantle
particles being scattered by the remnant and disrupting core.

\section{Discussion \& Conclusion}

\subsection{Summary}

As all the planets in the outer solar system have ring systems, we are
deeply interested in their formation and evolution. In this paper we
have investigated the formation of narrow rings through the tidal
disruption of a weak, gravitationally bound satellite that migrates
within its Roche limit. In our $N$-body numerical simulations, the
satellite is modelled as a rubble pile: a gravitational aggregate of
spherical particles that undergo inelastic collisions. We have
considered both homogeneous satellites and differentiated bodies, the
latter being composed of denser core particles surrounded by a less
dense mantle.

We have shown that the Roche limit for a rubble pile is closer to the
planet than for a fluid body of the same mean density, although its
location scales in the same way with the mass of the planet and the
mean density of the satellite. The Roche limit for a differentiated
body is also closer to the planet that for a homogeneous satellite of
the same mean density.

Within its Roche limit, a homogeneous satellite is totally disrupted
and forms a narrow ring.  The initial stages of the disruption are
similar to the evolution of a viscous fluid ellipsoid, which becomes
highly elongated in the azimuthal direction.  Later, gravitational
instability produces irregular structure and dynamics within the
proto-ringlet.

On the other hand, a differentiated satellite tends to undergo a
disruption of its mantle only. This process is similar to Roche-lobe
overflow in interacting binary stars, although proceeding
simultaneously through both Lagrange points L1 and L2, and produces
streams of ejected particles whose trajectories are initially well
described by solutions of the restricted three-body problem.  Again,
however, gravitational instability affects the streams near nodes
where the particles come almost to rest in the corotating frame and
the density is enhanced. This type of disruption process produces two
narrow rings on either side of a remnant satellite.
If a differentiated satellite is placed sufficiently close to the
planet, however, it undergoes a total disruption which has features of
both the processes described above.

In future work we will consider the evolution of the narrow rings and
attendant satellites formed by these mechanisms.

\subsection{Application to Specific Ring Systems}

Our results suggest that narrow rings such as those found around
Uranus could have formed by tidal disruption as opposed to the
generally favoured mechanism of disruption by external impact.
Depending on the properties of the original satellite, tidal
disruption can form either a single ring or two narrow rings with a
remnant satellite in between. Similarly, disruption of a number of
inwardly migrating satellites could produce a chain of narrow rings,
possibly interspersed with remnant satellites.

To date, shepherding by satellites is the only known mechanism of
maintaining the sharp edges of narrow rings as observed around Uranus
and Saturn. Only one of Uranus' rings (the
$\epsilon$ ring) has known shepherds, and even in this case there are
theoretical difficulties in accounting for the survival of the
configuration \citep{Goldreich:1987}.
If the rest of the rings in
the Uranian system have shepherds they must be small and dark. We
suggest that the elusive shepherds, if present, could be either the
undisrupted cores of differentiated satellites that suffered a mantle
disruption, or small undisrupted satellites that were either dense
enough or strong enough to resist tidal disruption.

Tidal disruption of a pre-existing satellite may be responsible for
the formation of some or all of the rings found around planets in our
solar system. The satellite must first migrate within its Roche limit;
this could come about through tidal interaction with the planet, if
the satellite orbits within the corotation radius of the planet, or
through planet-disk interaction in the early evolution of the system.
For the smaller ice giant planets Uranus and Neptune the Roche limit
(assuming a homogeneous fluid body with a density of
$1\,\mathrm{g}\,\mathrm{cm}^{-3}$) lies inside the corotation radius,
which means that satellites that were formed between these two radii
would migrate inwards through tidal interaction with the planet and
eventually be disrupted to form rings. This zone occupies roughly 67000--83000~km for Uranus and 71000-84000~km for Neptune. The process would be ongoing
until the reservoir of satellites was depleted, but is very slow
because of the long time-scale of tidally induced migration. These
successive tidal disruptions could explain the many narrow rings found
in the Uranian system if there was originally a significant reservoir
of small satellites between the Roche limit and the corotation
radius; there are currently eleven known moons of Uranus inside
corotation.

For the gas giant planets the Roche limit (calculated with the same
assumptions) is outside the corotation radius, which suggests that
tidal interaction with the planet could only lead to outward migration
and so could not be responsible for ring formation.  However, the
regular satellites of the gas giants are expected to form in a
circumplanetary gas disk. While the gas is present it could have
caused inward migration leading to ring formation through either
partial or total tidal disruption as suggested by \citet{Canup:2010}.
Indeed, inward type I migration is thought to be sufficiently rapid
that it may be difficult to explain the retention of any
proto-satellites while the gas remains. 
We note, however, that these comparisons of the corotation radius and Roche
radii are uncertain.  The corotation radius is expected to decrease
with time as a result of the contraction of the planet and the
reduction of its moment of inertia. Also, as we have shown, the Roche
radius for a rubble pile is smaller than for a fluid body, and the
Roche limit of a centrally condensed body is smaller than for a
homogeneous one.  Therefore it is conceivable the corotation radius
may have exceeded the appropriate Roche limit in the past \citep[as suggested by][]{Canup:2010} 
and that tidal migration may therefore have led to
ring formation around Jupiter and Saturn as well.

\section*{Acknowlegements}

The authors thank an anonymous referee for fast and constructive feedback. This research was supported by an STFC Rolling Grant and by a Royal Society Joint Project with Japan. ZML is currently supported by an STFC Advanced Fellowship.

\bibliography{/Users/zoe/Dropbox/Bibliography/GeneralBib}

\begin{thebibliography}{}

\bibitem[\protect\citeauthoryear{{Broadfoot}}{{Broadfoot}}{1986}]{Broadfoot:1986}
{Broadfoot} A.~L. et al.,  1986, Science, 233, 74

\bibitem[\protect\citeauthoryear{Burns, Hamilton \& Showalter}{Burns
  et~al.}{2001}]{Burns:2001}
Burns J.,  Hamilton D.~P.,    Showalter M.~R.,  2001, Dusty Rings and
  Circumplanetary Dust: Observations and Simple Physics.
Springer, pp 641--725

\bibitem[\protect\citeauthoryear{{Canup}}{{Canup}}{2010}]{Canup:2010}
{Canup} R.~M.,  2010, \nat, 468, 943

\bibitem[\protect\citeauthoryear{Chandrasekhar}{Chandrasekhar}{1961}]{Chandrasekhar:1961}
Chandrasekhar S.,  1961, Astrophysical Journal, 134, 662

\bibitem[\protect\citeauthoryear{{Chandrasekhar}}{{Chandrasekhar}}{1969}]{Chandrasekhar:1969}
{Chandrasekhar} S.,  1969, {Ellipsoidal figures of equilibrium}

\bibitem[\protect\citeauthoryear{Charnoz, Dones, Esposito, Estrada \&
  Hedman}{Charnoz et~al.}{2009}]{Charnoz:2009kx}
Charnoz S.,  Dones L.,  Esposito L.,  Estrada P.,    Hedman M.,  2009, in
  Dougherty M.,  Esposito L.,   Krimigis S.,  eds, Saturn form Cassini-Huygens
  Origin and evolution of saturn's ring system.
Springer, pp 537--575

\bibitem[\protect\citeauthoryear{{Charnoz}, {Morbidelli}, {Dones} \&
  {Salmon}}{{Charnoz} et~al.}{2009}]{Charnoz:2009}
{Charnoz} S.,  {Morbidelli} A.,  {Dones} L.,    {Salmon} J.,  2009, \icarus,
  199, 413

\bibitem[\protect\citeauthoryear{{Chau}, {Wong} \& {Wu}}{{Chau}
  et~al.}{2002}]{Chau:2002}
{Chau} K.~T.,  {Wong} R. H.~C.,    {Wu} J.~J.,  2002, {International Journal of
  Rock Mechanics and Mining Sciences}, 39, 69

\bibitem[\protect\citeauthoryear{{Christodoulou} \& {Narayan}}{{Christodoulou}
  \& {Narayan}}{1992}]{Christodoulou:1992}
{Christodoulou} D.~M.,  {Narayan} R.,  1992, \apj, 388, 451

\bibitem[\protect\citeauthoryear{Cook \& Franklin}{Cook \&
  Franklin}{1964}]{Cook:1964}
Cook A.~F.,  Franklin F.~A.,  1964, Astronomical Journal, 69, 173

\bibitem[\protect\citeauthoryear{{Dones}}{{Dones}}{1991}]{Dones:1991}
{Dones} L.,  1991, \icarus, 92, 194

\bibitem[\protect\citeauthoryear{{Esposito}}{{Esposito}}{2006}]{Esposito:2006}
{Esposito} L.,  2006, {Planetary Rings}

\bibitem[\protect\citeauthoryear{{Esposito}, {Brahic}, {Burns} \&
  {Marouf}}{{Esposito} et~al.}{1991}]{Esposito:1991}
{Esposito} L.~W.,  {Brahic} A.,  {Burns} J.~A.,    {Marouf} E.~A.,  1991,
  {Particle properties and processes in Uranus' rings}.
pp 410--465

\bibitem[\protect\citeauthoryear{{Goldreich} \& {Porco}}{{Goldreich} \&
  {Porco}}{1987}]{Goldreich:1987}
{Goldreich} P.,  {Porco} C.~C.,  1987, \aj, 93, 730

\bibitem[\protect\citeauthoryear{{Goldreich} \& {Soter}}{{Goldreich} \&
  {Soter}}{1966}]{Goldreich:1966}
{Goldreich} P.,  {Soter} S.,  1966, \icarus, 5, 375

\bibitem[\protect\citeauthoryear{{Goodman} \& {Narayan}}{{Goodman} \&
  {Narayan}}{1988}]{Goodman:1988}
{Goodman} J.,  {Narayan} R.,  1988, \mnras, 231, 97

\bibitem[\protect\citeauthoryear{{Harris}}{{Harris}}{1984}]{Harris:1984}
{Harris} A.,  1984, in {Greenberg} R.,  {Brahic} A.,  eds, {Planetary Rings}
  The origin and evolution of planetary rings.
University of Arizona Press, pp 641--659

\bibitem[\protect\citeauthoryear{{Higa}, {Arakawa} \& {Maeno}}{{Higa}
  et~al.}{1996}]{Higa:1996}
{Higa} M.,  {Arakawa} M.,    {Maeno} N.,  1996, \planss, 44, 917

\bibitem[\protect\citeauthoryear{{Higa}, {Arakawa} \& {Maeno}}{{Higa}
  et~al.}{1998}]{Higa:1998}
{Higa} M.,  {Arakawa} M.,    {Maeno} N.,  1998, \icarus, 133, 310

\bibitem[\protect\citeauthoryear{{Holsapple} \& {Michel}}{{Holsapple} \&
  {Michel}}{2006}]{Holsapple:2006}
{Holsapple} K.~A.,  {Michel} P.,  2006, \icarus, 183, 331

\bibitem[\protect\citeauthoryear{{Holsapple} \& {Michel}}{{Holsapple} \&
  {Michel}}{2008}]{Holsapple:2008}
{Holsapple} K.~A.,  {Michel} P.,  2008, \icarus, 193, 283

\bibitem[\protect\citeauthoryear{{Karkoschka}}{{Karkoschka}}{1997}]{Karkoschka:1997}
{Karkoschka} E.,  1997, \icarus, 125, 348

\bibitem[\protect\citeauthoryear{{Karkoschka}}{{Karkoschka}}{2001a}]{Karkoschka:2001a}
{Karkoschka} E.,  2001a, \icarus, 151, 51

\bibitem[\protect\citeauthoryear{{Karkoschka}}{{Karkoschka}}{2001b}]{Karkoschka:2001}
{Karkoschka} E.,  2001b, \icarus, 151, 78

\bibitem[\protect\citeauthoryear{Latter, Rein \& Ogilvie}{Latter
  et~al.}{2012}]{Latter:2012}
Latter H.,  Rein H.,    Ogilvie G.~I.,  2012, Monthly Notices of the Royal
  Astronomy

\bibitem[\protect\citeauthoryear{{Leinhardt} \& {Richardson}}{{Leinhardt} \&
  {Richardson}}{2002}]{Leinhardt:2002}
{Leinhardt} Z.~M.,  {Richardson} D.~C.,  2002, \icarus, 159, 306

\bibitem[\protect\citeauthoryear{{Leinhardt}, {Richardson} \&
  {Quinn}}{{Leinhardt} et~al.}{2000}]{Leinhardt:2000}
{Leinhardt} Z.~M.,  {Richardson} D.~C.,    {Quinn} T.,  2000, \icarus, 146, 133

\bibitem[\protect\citeauthoryear{{Leinhardt} \& {Stewart}}{{Leinhardt} \&
  {Stewart}}{2012}]{Leinhardt:2012}
{Leinhardt} Z.~M.,  {Stewart} S.~T.,  2012, \apj, 745, 79

\bibitem[\protect\citeauthoryear{{Ockert}, {Cuzzi}, {Porco} \&
  {Johnson}}{{Ockert} et~al.}{1987}]{Ockert:1987}
{Ockert} M.~E.,  {Cuzzi} J.~N.,  {Porco} C.~C.,    {Johnson} T.~V.,  1987,
  \jgr, 92, 14969

\bibitem[\protect\citeauthoryear{{Ogilvie} \& {Lin}}{{Ogilvie} \&
  {Lin}}{2004}]{Ogilvie:2004}
{Ogilvie} G.~I.,  {Lin} D.~N.~C.,  2004, \apj, 610, 477

\bibitem[\protect\citeauthoryear{{Papaloizou} \& {Lin}}{{Papaloizou} \&
  {Lin}}{1989}]{Papaloizou:1989}
{Papaloizou} J.~C.~B.,  {Lin} D.~N.~C.,  1989, \apj, 344, 645

\bibitem[\protect\citeauthoryear{{Pollack}}{{Pollack}}{1975}]{Pollack:1975}
{Pollack} J.~B.,  1975, \ssr, 18, 3

\bibitem[\protect\citeauthoryear{{Porco}}{{Porco}}{1990}]{Porco:1990}
{Porco} C.~C.,  1990, Advances in Space Research, 10, 221

\bibitem[\protect\citeauthoryear{{Richardson}}{{Richardson}}{1994}]{Richardson:1994}
{Richardson} D.~C.,  1994, \mnras, 269, 493

\bibitem[\protect\citeauthoryear{{Richardson}, {Bottke} \& {Love}}{{Richardson}
  et~al.}{1998}]{Richardson:1998}
{Richardson} D.~C.,  {Bottke} W.~F.,    {Love} S.~G.,  1998, \icarus, 134, 47

\bibitem[\protect\citeauthoryear{{Richardson}, {Leinhardt}, {Melosh}, {Bottke}
  Jr. \& {Asphaug}}{{Richardson} et~al.}{2002}]{Richardson:2002}
{Richardson} D.~C.,  {Leinhardt} Z.~M.,  {Melosh} H.~J.,  {Bottke} Jr. W.~F.,
   {Asphaug} E.,  2002, Asteroids III, pp 501--515

\bibitem[\protect\citeauthoryear{{Showalter}}{{Showalter}}{1993}]{Showalter:1993}
{Showalter} M.~R.,  1993, in AAS/Division for Planetary Sciences Meeting
  Abstracts \#25 Vol.~25 of Bulletin of the American Astronomical Society,
  {Longitudinal Variations in the Uranian Lambda Ring}.
p.~1109

\bibitem[\protect\citeauthoryear{{Showalter} \& {Lissauer}}{{Showalter} \&
  {Lissauer}}{2006}]{Showalter:2006}
{Showalter} M.~R.,  {Lissauer} J.~J.,  2006, Science, 311, 973

\bibitem[\protect\citeauthoryear{{Sridhar} \& {Tremaine}}{{Sridhar} \&
  {Tremaine}}{1992}]{Sridhar:1992}
{Sridhar} S.,  {Tremaine} S.,  1992, \icarus, 95, 86

\bibitem[\protect\citeauthoryear{{Stadel}}{{Stadel}}{2001}]{Stadel:2001}
{Stadel} J.~G.,  2001, PhD thesis, UNIVERSITY OF WASHINGTON

\bibitem[\protect\citeauthoryear{{Tittemore} \& {Wisdom}}{{Tittemore} \&
  {Wisdom}}{1990}]{Tittemore:1990}
{Tittemore} W.~C.,  {Wisdom} J.,  1990, \icarus, 85, 394

\end{thebibliography}

\end{document}